\documentclass[iop]{emulateapj}
\shorttitle{Dwarfs and Tidal Features Around NGC\,7331}
\shortauthors{Ludwig et al.}
\usepackage[dvips]{color}

\begin{document}
\submitted{submitted: 2012 March 14; accepted: 2012 September 26}
   \title{
Giant Galaxies, Dwarfs, and Debris Survey. I. Dwarf Galaxies and Tidal
Features Around NGC\,7331}

\author{Johannes Ludwig, Anna Pasquali, Eva K.\ Grebel}

\email{ludwig@ari.uni-heidelberg.de}

\affil{Astronomisches Rechen-Institut, Zentrum f\"ur Astronomie der 
Universit\"at Heidelberg, M\"onchhofstr.\ 12--14, D-69120 Heidelberg, Germany}\

\and

\author{John S.\ Gallagher III}

\affil{Department of Astronomy, University of Wisconsin, 475 North Charter Street, Madison, WI 53706-1582, USA}\

\begin{abstract}
The Giant GAlaxies, Dwarfs, and Debris Survey concentrates on
the nearby universe to study how galaxies have interacted in groups of
different morphology, density, and richness.  In these groups we
select the dominant spiral galaxy and search its surroundings for
dwarf galaxies and tidal interactions.  This paper presents the first
results from deep wide-field imaging of NGC\,7331, where we detect
only four low luminosity candidate dwarf companions and a stellar
stream that may be evidence of a past tidal interaction.  The dwarf
galaxy candidates have surface brightnesses of  $\mu_{r}  \approx$
23--25 mag~arcsec$^{-2}$ with ($g-r$)$_0$ colors of 0.57--0.75~mag in
the Sloan Digital Sky Survey filter system, consistent with their
being dwarf spheroidal galaxies (dSph).  A faint stellar stream structure on
the western edge of NGC\,7331 has $\mu_g \approx$ 27 mag~arcsec$^{-2}$
and a relatively blue color of ($g-r$)$_0$ = 0.15~mag. If it is tidal
debris, then this stream could have 
formed from a rare type of
interaction between NGC\,7331 and a dwarf irregular or transition-type
dwarf galaxy.  We compare the structure and local environments of
NGC\,7331 to those of other nearby giant spirals in small galaxy
groups.  NGC\,7331 has a much lower ($\sim$2\%) stellar mass in the
form of early-type satellites than found for M31 and lacks the
presence of nearby companions like luminous dwarf elliptical galaxies or the
Magellanic Clouds. However, our detection of a few dSph candidates
suggests that it is not deficient in low-luminosity satellites.
\end{abstract}

\keywords{galaxies: dwarf – galaxies: evolution – galaxies: formation
galaxies: groups: individual: NGC\,7331, galaxies: interactions, galaxies: structure}

\section{introduction}
\label{introgg}

Current models of galaxy formation and evolution suggest that the main
process of mass assembly is hierarchical growth via minor and major
mergers \citep[e.g.,][]{white1991ApJ}.  Observationally this
hierarchical structure formation can be traced by the characteristic
large-scale galaxy distribution of clusters of galaxies as the most
luminous and massive concentrations of matter.  Most of the luminous
mass, however, is not located in these high-density knots of the
cosmic web, but instead along its numerous filaments in less massive
galaxy groups.  In fact, about 85\% of the nearby galaxies are not
found in galaxy clusters, but are located instead in galaxy groups and
in the field \citep{Tully1987ApJ321,Karachentsev2005AJ129}.
\citet{Eke2005MNRAS362} found that the fraction of mass in stars
increases with decreasing group size, with Local-Group-sized galaxy
agglomerations containing most of the stellar mass.  

There are many indications that mergers played a role in forming
groups. For instance, the clustering strength increases with group
luminosity \citep{Padilla2004MNRAS352}.  Within the groups themselves
the earlier-type galaxies are more strongly clustered and are usually
located closer to the group center \citep{Girardi2003AAp406},
mimicking the morphology-density relation of galaxy clusters
\citep[e.g.,][]{Oemler1974ApJ194,Dressler1980ApJ236}.  Dwarf galaxies
and satellites also exhibit a morphology-distance and
morphology-density relation with respect to their location within
groups
\citep[e.g.,][]{Einasto1974Nature252,vdBergh1994AJ107,Grebel1997RevModAst10}.
Galaxy groups range from poor to rich groups and from apparently
little evolved diffuse ``clouds'' to compact groups with past or
ongoing major mergers.  This range of types allows us to explore
galaxy evolution and environmental effects in groups of different
density, and to compare their properties and evolutionary state with those
of galaxies in the field or in galaxy clusters
\citep{Grebel2007ESOSymp}.  The evolutionary end stage may be
represented by the so-called fossil groups that contain a dominant,
large elliptical galaxy in their center, likely the result of 
mergers \citep{Jones2003MNRAS343,DOnghia2005ApJL630}.

The best-studied group of galaxies is obviously the Local Group.  Its two
dominant spiral galaxies, the Milky Way (MW) and M31, are surrounded by a
large number of mainly early-type dwarfs, many of which were only
detected in recent years
\citep[e.g.,][]{Zucker2004ApJL612,Zucker2006ApJL650,Zucker2007ApJ659,Belokurov2006ApJL647,Belokurov2007ApJ654,Martin2009ApJ705,Richardson2011ApJ732}.
The vast majority of the Local Group dwarfs are dwarf spheroidal
(dSph) galaxies.  The gas-deficient dSphs as well as dwarf ellipticals
(dEs) are typically found in close proximity to one of the large
spirals, whereas most of the gas-rich, star-forming dwarf irregulars
(dIrrs) are located at larger distances.  For a review of the
properties of different types of dwarf galaxies, see
\citet{Grebel2001ApSSS277}. Evidence for minor merger events, either
in the form of ongoing interactions or relics in the form of vast
tidal streams, is found in and around the MW
\citep[e.g.,][]{Ibata1994Nature370,Yanny2003ApJ588,Duffau2006ApJL636,Bell2008ApJ680,Williams2011ApJ728}
and M31
\citep[e.g.,][]{Ibata2001Nature412,Zucker2004ApJL612,Chapman2008MNRAS390,McConnachie2009Nature461,Mackey2010ApJL717}.

Other nearby groups also show structures dominated by two or more
moderately massive galaxies with an extended entourage of dwarfs
\citep[e.g.,][]{Karachentsev2002AAp385,Karachentsev2002AAp383,Karachentsev2003AAp408,Karachentsev2003AAp404,Karachentsev2003AAp398,Chiboucas2009AJ137}.
For galaxy groups beyond 5 Mpc, usually less detailed information is
available, with past efforts having focused mainly on global
properties
\citep[e.g.,][]{Trentham2006MNRAS369,Tully2008AJ135,Jacobs2009AJ138,Courtois2009AJ138}.
Furthermore, prominent tidal streams and morphological perturbations
were detected around individual nearby spiral and elliptical galaxies
\citep{Schweizer1988ApJ,Malin1999ASPC182,MartinezDelgado2009ApJ692,MartinezDelgado2010AJ140,Mouhcine2010ApJ714,Mouhcine2011MNRAS415,Mouhcine2009MNRAS399,Miskolczi2011A&A536}
or on larger scales in, e.g., the interacting M81 group
\citep[e.g.,][]{Yun1994Nature372,Makarova2002AAp396}.

The galaxy content at the faint end of the luminosity function and the
presence or absence of tidal streams and substructure allow us to
visualize different steps of group evolution driven by the environment
similar to related studies in galaxy clusters
\citep[e.g.,][]{Lisker2006AJ132,Lisker2007ApJ660,Smith2009MNRAS392,Paudel2010MNRAS405}.
This motivated us to initiate the ``Giant GAlaxies, Dwarfs, and
Debris Survey'' (GGADDS).  Our goals are to correlate the observed
properties of spiral galaxies and groups to the frequency of tidal
interactions, and to characterize the impact of local density on the
dwarf galaxy population.  GGADDS combines the search for signatures of
recent accretion events with the study of dwarf galaxy populations in
different types of nearby galaxy groups.

In our current paper, we present the first GGADDS results based on deep
imaging of the SAb spiral galaxy NGC\,7331 and its surroundings.  In
Section~\ref{project}, we introduce the GGADDS project.  In
Section~\ref{obsgg}, we describe the observations and data reduction
for our first target, NGC\,7331.  In Section~\ref{results}, we present
our results, followed by a discussion (Section~\ref{discussion}) and
our conclusions (Section~\ref{conclusions}).

\section{The GGADDS project}
  \label{project}

The GGADDS project concentrates on galaxy groups in the nearby  
universe within a distance range of $\sim 10$ to $\sim 35$~Mpc.  
The main goal is to detect dwarf galaxy candidates and
possible tidal features in order to constrain their properties
as a function of environment, and to explore whether links exist 
between these two aspects of the surroundings of giant galaxies. 
We currently use deep imaging with small to
medium-size optical telescopes (0.9m to 4m-class) to 
support photometric and structural studies, later to be
complemented with spectroscopy.  Our selected dominant group galaxies
 range from $M_B = -19$ to $-23$~mag and are located in
environments with densities varying from 0.08 to 1.6
galaxies~Mpc$^{-3}$.  

The distance range of our targets is motivated
by our choice of telescopes: at 10 -- 35~Mpc,
the instrument field of view of commonly used wide-field
imagers ($\sim36' \times 36'$ to $59' \times 59'$) covers an area of typically 
($175 \times 175$)~kpc$^{2}$ 
to  ($611 \times 611$)~kpc$^{2}$, 
while the instrument sensitivity 
allows us to reach  a  limiting surface brightness of 27 mag~arcsec$^{-2}$
in the $g$-band. For comparison, numerical simulations by  \citet{Johnston2008ApJ} show that tidal streams
due to accretion events should be detectable at this surface
brightness 
for minor mergers that occurred in the
last Gyr. Minor mergers have baryonic mass ratios $\leq 1:4$ for $M_{satellite}/M_{primary}$ \citep{Lotz2011ApJ742}.

To test the feasibility of this project we observed NGC\,7331, an SAb
galaxy at a distance of 14.2~Mpc viewed under an inclination angle
$i=77\degr$. Its inferred luminosity is $M_B = -20.4$~mag
\citep{deVaucouleurs1991trcb.book} and its maximum rotational velocity
is ($245.5\pm5.2$)~km~s$^{-1}$ \citep[taken from
Hyperleda\footnote{http://leda.univ-lyon1.fr/}, based on the catalog
from][]{Bottinelli1982A&AS47}. These quantities   are similar to
the properties of the MW and M31.  NGC\,7331 contains a large-scale
dust ring with a radius of about 6 kpc \citep{Regan2004ApJS} and its
observed H\,{\sc i} distribution shows  a warp in the outskirts of the
disk \citep{Bosma1981AJ86}.  The latter could be a sign of past
interactions.

\begin{figure*}[b]
\includegraphics[width=7.1in]{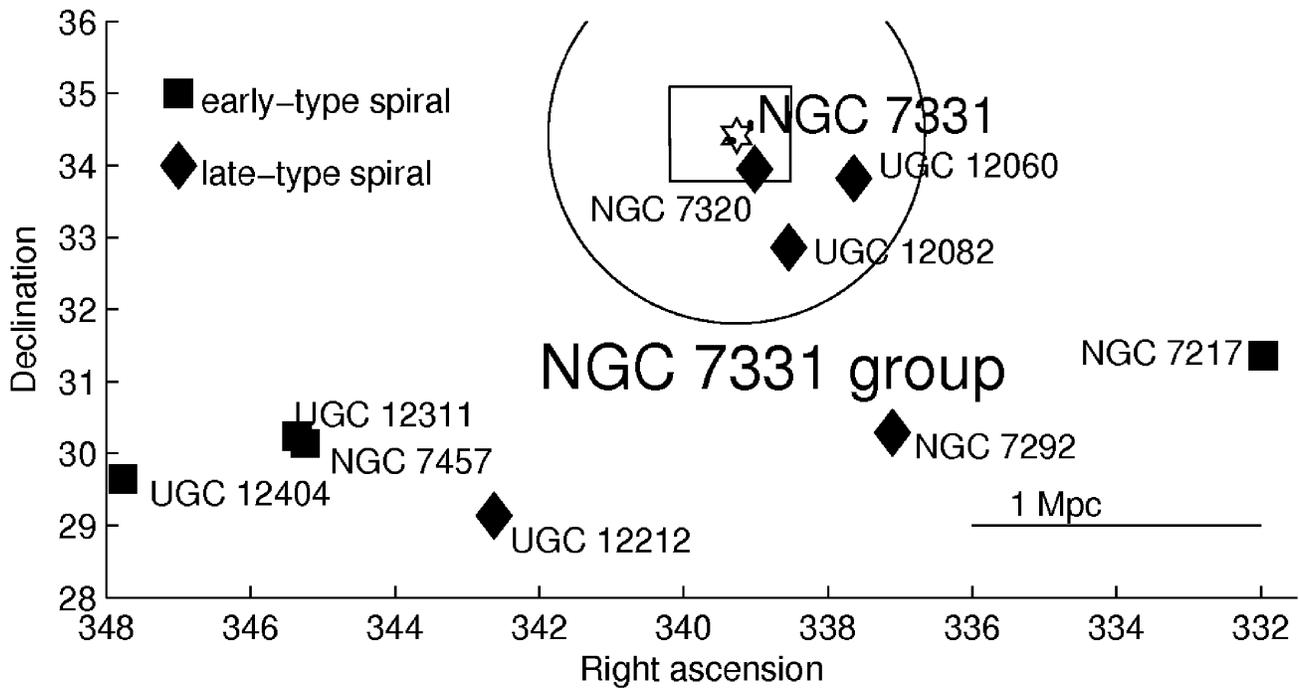}

\caption{Wider NGC\,7331 group environment.  The star symbol
represents the centroid of NGC\,7331.  The dots around it are our dwarf
galaxy candidates. The open rectangle marks our field of view.
Filled squares show other massive, yet less luminous early-type 
spiral galaxies in the NGC\,7331 group and in its immediate surroundings. 
Filled lozenges show late-type spirals in the group and in its vicinity.
The open circle indicates a virial radius of 1.3~Mpc around NGC\,7331.}
\label{7331group}       
\end{figure*}

NGC\,7331 is located in a sparse group with a density of 0.33 galaxies
Mpc$^{-3}$ down to $M_{V}=-16$~mag \citep{Tully1988}.  Other luminous
group members include the disk galaxies NGC\,7217, NGC\,7320,
NGC\,7292, NGC\,7457, UGC\,12060, UGC\,12082, UGC\,12212, UGC\,12311,
and UGC\,12404 (see Figure \ref{7331group}).  The galaxies of this
group are within $\pm 170$~km~s$^{-1}$ from NGC\,7331, which itself
has a radial velocity $ v=(816 \pm 1)$~km~s$^{-1}$
\citep[]{Haynes1998AJ}. NGC\,7331 is the most luminous galaxy of this
group.  The other late-type galaxies span the magnitude range  $-16
\lesssim M_{B}$ $\lesssim -20$~mag. In comparison with the Local
Group the NGC\,7331 group has only one member brighter than
$M_{B}=-20$~mag.  The NGC\,7331 group is classified as a stable group
\citep{Materne1974A&A35} and is  part of the Pegasus Spur
\citep{Tully1988}.

Figure \ref{7331group} gives an overview of the NGC\,7331 galaxy group, 
underlining the loose, extended character of this group, which resembles the nearby Sculptor group \citep[e.g., ][]{Karachentsev2003AAp404}.
When measuring the projected distances from NGC\,7331 to the other members, 
we get typical distances of 1~Mpc and for the 
most distant member, UGC 12404, 2.16~Mpc.

\begin{figure*}[b]
\includegraphics[width=7.1in]{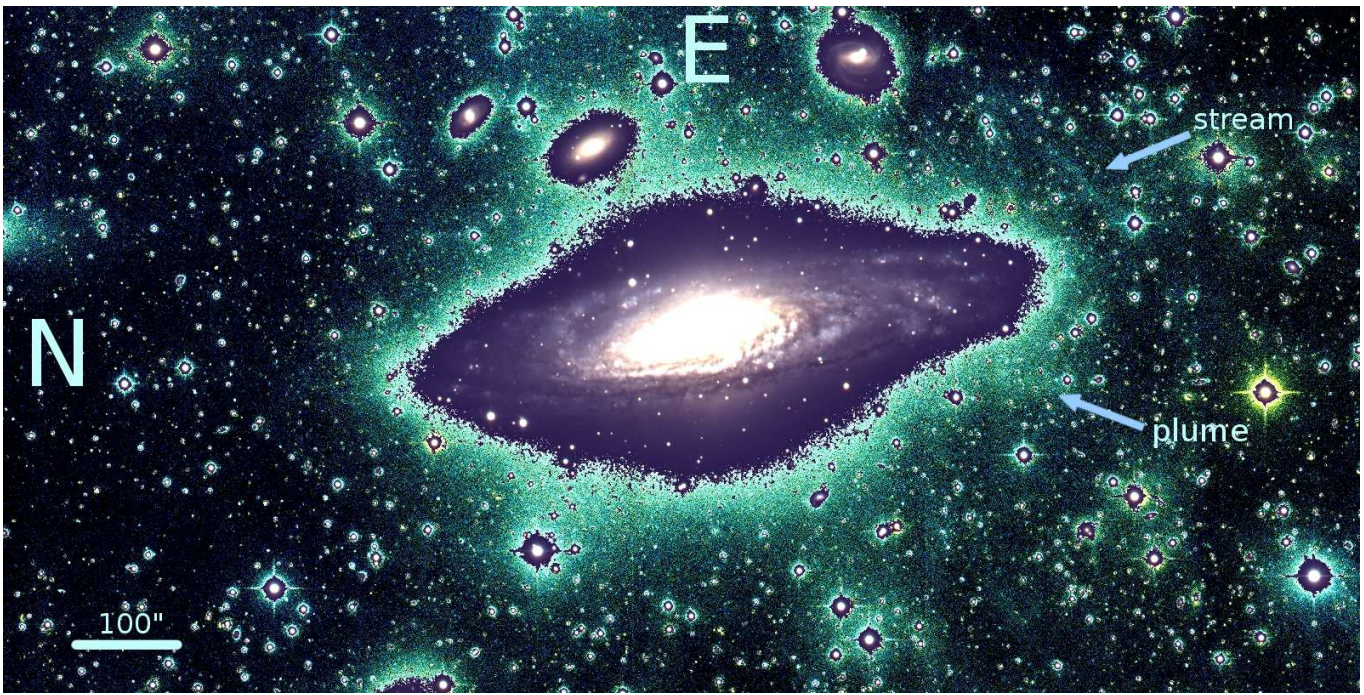}
\caption{Optical view of the inclined spiral galaxy NGC\,7331
(center).  The image is composed of deep images in the SDSS $g$- and
$r$-bands obtained with the 0.9m WIYN telescope and has a total size 
of $22.5' \times 11.4'$, where 100$''$ correspond to 7 kpc.  A number of luminous
background galaxies and foreground stars are visible as well.  The
outer, grainy-looking regions are contrast enhanced in order to make 
faint features visible.  In the outskirts, faint tidal debris structure 
can be seen and is highlighted by the arrows.}
 
\label{7331hc}       
\end{figure*}

\section{Observations and data reduction}
\label{obsgg}

The observations were performed in 2009 October with the Wisconsin
Indiana Yale NOAO (WIYN) 0.9m telescope at Kitt Peak National
Observatory in Arizona equipped with the MOSAIC imager. This
instrument is a wide-field imager with eight single CCDs, providing a
total area of ($8192 \times 8192$) pixels$^2$. The total field of view
is 1 deg$^{2}$ with a pixel scale of 0.43$''$~pixel$^{-1}$.

We imaged NGC\,7331 for 3.5 hr (9 $\times$ 1200~s + 2 $\times$ 900~s)
in the Sloan Digital Sky Survey (SDSS) $g$-band and 4.17 hr (11
$\times$ 1200~s + 2 $\times$ 900~s) in the SDSS $r$-band.  The dither
pattern was chosen with a step size of 400$''$ in right ascension and
declination so that the galaxy would lie in each of the central four
CCDs.

We reduced the data with IRAF\footnote{IRAF is distributed by the National Optical Astronomy
Observatory, which is operated by the Association of Universities for
Research in Astronomy, Inc., under cooperative agreement with the
National Science Foundation.} using the MSCRED package, which was
specifically developed for the MOSAIC imager.
As a first step, the images were corrected for bias and dome
flatfields in order to remove the 
pixel-to-pixel sensitivity variations. We also corrected for the illumination pattern
using a \textquotedblleft super flat\textquotedblright, which was produced by taking the median of the science exposures 
(making sure to remove the left-over sources), as well as for atmospheric
extinction. The images were scaled to the same sky transparency.

We estimated the accuracy of the flat-field correction by comparing the 
global mode of the background (computed across the overall field of view) 
with local background modes (calculated in 16 subarrays, each half of the size
of a CCD). 
We measured a typical variation of the background values of about
1.7\% in the SDSS $r$-band and of 2.3\% in the $g$-band.

We used stars from the USNO-B1 catalog \citep{Monet2003AJ125} to derive a world coordinate system (WCS) solution for each CCD in each exposure. 
The typical accuracy of the transformation from pixel coordinates to equatorial coordinates is
0.2$''$ -- 0.3$''$ across the 1 deg$^{2}$ field of view. 
We then ran the IRAF task {\sf mscimage} to combine all eight CCDs of one exposure into a single mosaic image, reproducing the full field of view
of the instrument with a fixed pixel scale of 0.43$''$~pixel$^{-1}$.

Individual mosaic images were subsequently corrected for the airmass with the 
extinction coefficients provided by the observatory\footnote{The standard extinction values for the Kitt Peak National Observatory can be found
  in the file {\sf kpnoextinct.dat} of the task {\sf onedstds}, which is part of the IRAF package ONEDSPEC} 
at the central wavelengths of the filters\footnote{http://www.noao.edu/kpno/mosaic/filters/filters.html}.

For each individual image we calculated the mode value of the background across the entire field of view and subtracted it. 
Furthermore, in order to account for the different transparency of the atmosphere in different
nights, we measured the total counts in
a non-saturated region centered on NGC\,7331. We chose
the exposure with the highest counts as reference and scaled the others to this value.
We stacked the images by using their WCS solution and normalized the final image
by the total exposure map. 
Due to the dither pattern we obtain a final image size of 11148 pixels $\times$ 11756 pixels with
a resulting field of view of 
$1.3\degr \times 1.4\degr$, where the edges of the final image have a lower total 
exposure time (fewer exposures) and a higher noise than the central regions.
The area with complete overlap of all stacks is 5440 pixels $\times$ 4640 pixels resulting in a field of view of 33$' \times$ 38$'$.
The final image is characterized by an average point-spread
function (PSF) with a full width at half maximum (FWHM) of 1.3$''$ in each filter.

For the photometric calibration we used stars in common between our
final image and the SDSS Data Release 7
\citep[DR7;][]{Abazajian2009ApJS182}. We only used stars that are not
saturated in our science images but bright enough to have a good
signal-to-noise ratio (S/N). In the SDSS $g$-band we used stars
between 16.5 and 18~mag and derived a zero point of (22.24 $\pm$
0.04)~mag, while in the SDSS $r$-band we selected stars between 16 and
17.5~mag and obtained a zero point of (22.19 $\pm$  0.03)~mag.

We used the extinction maps of \citet{Schlegel1998ApJ500} and the
extinction law of \citet{Cardelli1989ApJ345} with $R_{V}=3.1$ to
calculate the Galactic foreground extinction at the central wavelength
of the SDSS $g$- and $r$-bands ($R_{V}$ is the ratio of total
to selective extinction in the $V$-band). All the magnitudes and
surface brightnesses derived in the next sections are given already
corrected for Galactic foreground extinction.

In deep imaging Galactic cirrus can act as an intervening contaminant. 
Therefore, the occurrence of cirrus imposes 
a limit on surface photometry as it is a Galactic foreground that cannot 
be removed \citep[e.g.,][]{Sandage1976AJ81}.

In our observed region Galactic cirri are visible. Interestingly
cirrus clouds show very red broadband colors
\citep{Szomoru1998ApJ494}. Typical $(B-R)$ values range from 1.0 to
1.7~mag \citep{Guhathakurta1994ASPC58} and translate into $(g-r)=1.33$
to $2.03$~mag  when using the color transformation
$(B_{J}-R)=(g-r)+0.33$ from \citet{Fukugita1995PASP107}.  Therefore,
on the basis of their measured colors, faint and blue galaxy features
can be distinguished from red Galactic cirrus clouds.
 
The extinction maps of \citet{Schlegel1998ApJ500} do not take into account 
the presence of small scale dust features which can be seen in emission as 
``cirrus".  Therefore our reddening correction does not remove any effect 
associated with source obscuration by dust associated with cirrus.   
A careful visual inspection of our images indicates that no cirrus is in 
close proximity to the objects discussed below, so cirrus reddening is 
unlikely to be a significant effect. 

Observationally, the angular resolution is the most limiting factor for detecting faint, distant dwarf galaxies  
or streams. Due to the pixel scale of 0.43$''$~pixel$^{-1} $ and the need to have several pixels
in order to resolve structures we are able to detect objects with a minimum size of 0.3 kpc at the distance of NGC\,7331.

\section{Results}
\label{results}
\subsection{NGC\,7331}

A color composite image of NGC\,7331 is shown in Figure \ref{7331hc}.
Two images with different contrast levels are overlaid. The white-blue
high-surface-brightness inset shows the brighter regions of NGC\,7331,
highlighting its overall structure, especially its spiral arms and
bulge.  The underlying green low S/N image reveals the
low-surface-brightness features of the galaxy, including a candidate
tidal stream and a plume, which are marked with arrows in Figure
\ref{7331hc} and are discussed further below.

\subsection{Identification of Dwarf Galaxy Candidates}

In order to identify possible dwarf galaxy candidates around
NGC\,7331, a first analysis was done via visual inspection of the
science images.  We selected dwarf galaxy candidates on the basis of
their low apparent surface brightness, diffuse structure, and angular
extent of at least 7$''$, which is five times the typical PSF. 
This was necessary in order to be able to distinguish
extended objects from point sources.  With spectroscopic follow-up
studies \citet{Chiboucas2010ApJ} confirm that these selection criteria
have a good success rate in identifying true dwarf galaxies.  Our
field of view and sensitivity do in principle permit us to detect all
objects within a galactocentric distance of up to 170~kpc around
NGC\,7331 and down to a surface brightness $\mu_{g}=
27$~mag~arcsec$^{2}$.

We identified four candidate dwarf galaxies, which we named with
capital Latin letters in alphabetical order according to increasing
projected distance to NGC 7331.  Thus the nearest one is called
NGC\,7331 A, the second nearest NGC\,7331 B, and so forth.

\begin{figure}[hb]
\includegraphics[width=3.425in]{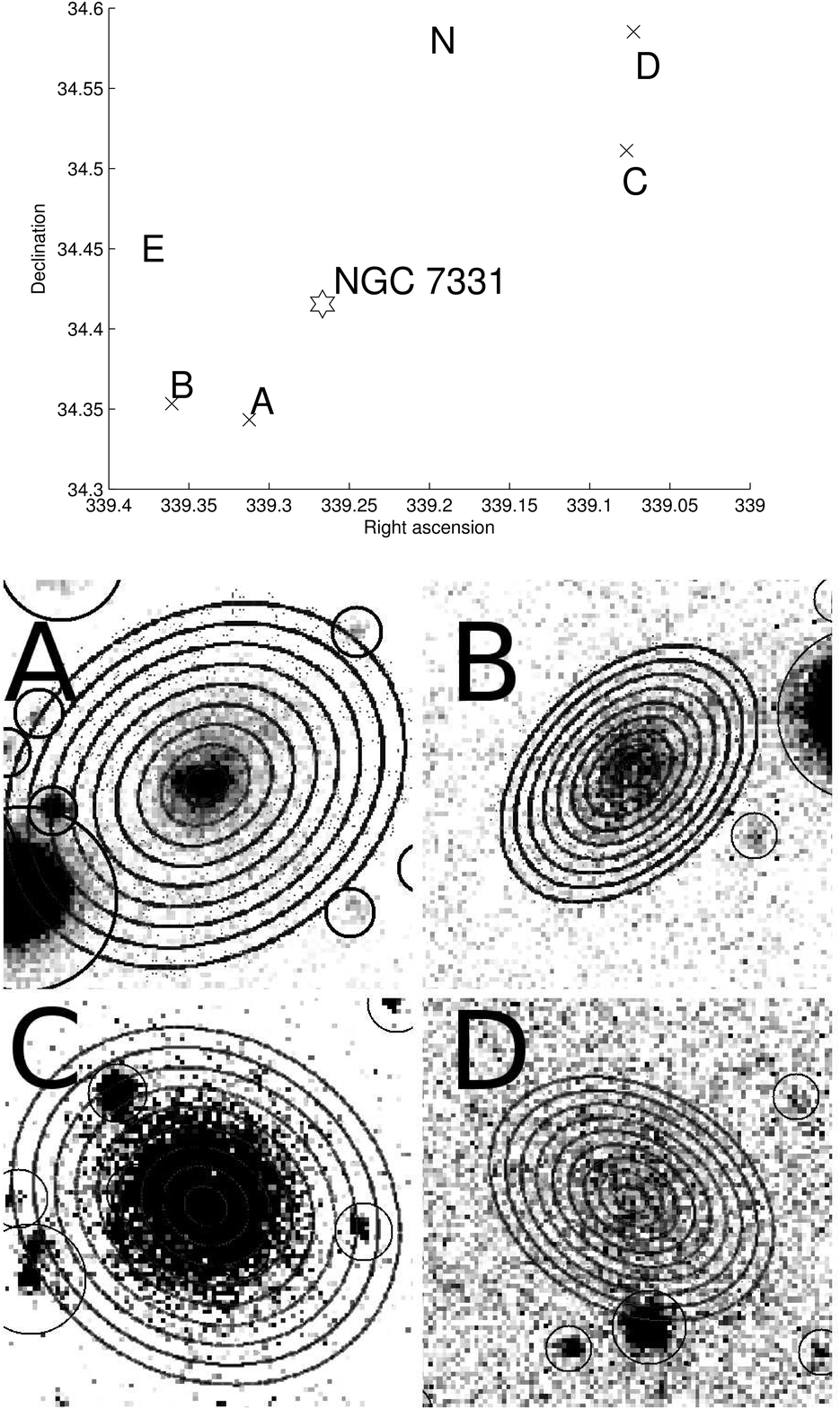}
\caption{Top: the positions of the dwarf galaxy candidates.
Bottom:
images of the dwarf galaxy candidates in the SDSS $g$-band obtained
with the WIYN 0.9m telescope. These
are cut-outs of the final stacked image of our observations
with sizes of 180 pixels $\times$ 180 pixels. North is left and east is up.
The concentric elliptical annuli centered on each of the dSph candidates indicate a subset of the regions
within which the surface brightness was measured.  The open circles mark presumed
background objects that were masked out.}
\label{7331g_dgal}       
\end{figure}

\subsection{Photometry of the Dwarf Galaxy Candidates}
\label{photdgal}

We ran SExtractor \citep{Bertin1996A&AS117} on images with a size of
180 $\times$ 180 pixels centered on each candidate in order to measure
the integrated $g$ and $r$ magnitudes of the candidates after an
accurate estimate of their local background.  We configured SExtractor
to measure isophotal magnitudes in two different apertures defined by
the detection thresholds of 1$\sigma$ and 3$\sigma$ above the background.
The resulting apparent integrated magnitudes are given in Table
\ref{dwarfdata}, together with their errors computed from SExtractor,
which also takes into account the uncertainty on the photometric
calibration. The absolute magnitudes were calculated assuming the same
distance modulus (30.70 $\pm$ 0.32~mag) for the dwarf galaxy
candidates as for NGC\,7331. Their associated errors also include the
error in the distance modulus. 

We see that these four candidates fall in the magnitude range
$-13\lesssim M_{r} \lesssim -11$ typical of the classical dSph galaxies in the
Local Group \citep{Grebel2003AJ125}. We checked the H\,{\sc i} data
obtained for NGC\,7331 by \citet{Walter2008AJ136}, and found no
H\,{\sc i} emission at the position of the four candidates, indicating
that there may be no gas for future star formation available.
This would be consistent with the properties of the majority of known
dSphs in the Local Group.  Only NGC\,7331 A could possibly be
classified as a transition object, because its core, being slightly bluer
than its outer parts (see Section \ref{surfacebrightnessprofiles}), 
could have recently undergone some star formation.

As the S/N is different for the $g$- and $r$-bands, 
the integrated magnitudes computed by SExtractor sample different areas within each galaxy candidate. 
In the next section, we will use the IRAF task {\sf ellipse} to compute the integrated color 
over the same region for each candidate and filter.

\begin{table*}[t]
 \centering
\caption{Observational Properties of the Dwarf Galaxy Candidates}
\label{dwarfdata}
\begin{tabular}{p{5cm}llll}

Property &  NGC\,7331 A & NGC\,7331 B & NGC\,7331 C & NGC\,7331 D\\ 

\hline
\hline

R.A. [J2000] [h m s] (1)  & 22:37:14.98 & 22:37:26.56 & 22:36:18.49 & 22:36:17.46 \\

Decl. [J2000] [$^{\circ} $ m s] (2)  & 34:20:35.86  & 34:21:12.23 & 34:30:40.37 & 34:35:07.20  \\

 $g_{0}$  $(1\sigma)$ (mag) (3) &  18.48 $\pm$ 0.04 &  20.01 $\pm$ 0.05 & 18.30 $\pm$ 0.04 & 20.65 $\pm$ 0.05 \\

 $g_{0}$  $(3\sigma)$ (mag) (4) & 19.91 $\pm$ 0.04 &  20.82 $\pm$ 0.20 & 18.53 $\pm$ 0.04 & --- \\ 

 $r_{0}$  $(1\sigma)$ (mag) (5) &  17.80 $\pm$ 0.03 &  18.75 $\pm$ 0.03 & 17.58 $\pm$ 0.03 & 20.21 $\pm$ 0.04 \\

 $r_{0}$  $(3\sigma)$ (mag) (6) &  19.14 $\pm$ 0.03 &  19.75 $\pm$ 0.03 & 17.90 $\pm$ 0.03 & --- \\

$A_{g}$ (mag) (7) &   0.34 & 0.32 & 0.33 & 0.33   \\

$A_{r}$ (mag) (8) &   0.25 & 0.23 & 0.24 & 0.24   \\

$(g-r)_{0}$ $(1\sigma)$ (mag) (9) & 0.57 $\pm$ 0.05 & 0.63 $\pm$ 0.06 & 0.55 $\pm$ 0.05 & 0.75 $\pm$ 0.06    \\

$(g-r)_{0}$ $(3\sigma)$ (mag) (9) & 0.39 $\pm$ 0.05 & 0.62 $\pm$ 0.20 & 0.54 $\pm$ 0.05 &  ---  \\     

$\mu_{0}(r)$ (mag~arcsec$^{-2}$) (10) & 23.33  $\pm$ 0.10 & 24.10  $\pm$ 0.06 & 23.76  $\pm$ 0.03 & 25.20 $\pm$ 0.08          \\

$h_{r}$ (pc) (11) & 325  $\pm$ 33 & 456 $\pm$ 15 & 592  $\pm$ 16 & 516 $\pm$ 27                          \\

$n $    (12)  & 1.28 $\pm$ 0.11 & 0.62 $\pm$ 0.07 & 0.71 $\pm$ 0.03 & 0.52 $\pm$ 0.07                      \\

$M_{r}$ $(1\sigma)$ (mag) (13) & -12.90 $\pm$ 0.32 & -11.95 $\pm$ 0.32 & -13.12 $\pm$ 0.32 & -10.48 $\pm$ 0.32    \\

$R$\,(NGC\,7331)$_{\rm projected}$ (kpc) (14) &  20 $\pm$ 2 & 25 $\pm$ 3 & 45 $\pm$ 5 & 58 $\pm$ 7 \\

\hline

\end{tabular}

Rows (1) and (2) are the coordinates (J2000 right ascension and
declination) where we give hours, minutes, and seconds in row (1) and
degrees, minutes, and seconds in row (2). Rows (3) -- (6) are the
apparent magnitudes in the SDSS $g$- and $r$-band with 1$\sigma$ and
3$\sigma$ thresholds above the background detections. In rows (7) and
(8),  $A_{g}$ and $A_{r}$ are the extinction values for the SDSS $g$-
and $r$-band. Row (9) contains the extinction-corrected color in the
SDSS $g$- and $r$-bands. Row (10) shows the central surface brightness
in the $r$-band, $\mu_{0}(r)$. Row (11) lists the exponential scale
length, $h_{r}$.  Row (12) provides the S\'ersic index; $n$, row (13)
the absolute magnitude in the $r$-band, $M_{r}$, when assuming a
distance modulus of 30.70$\pm$ 0.32; and row (14) the projected
distance to NGC\,7331, $R$.

\end{table*}

\subsection{Surface Brightness Profiles}
\label{surfacebrightnessprofiles}

At first we measured the photometric center of our four dwarf
candidates with the IRAF tool {\sf center} in the DIGIPHOT package.
The resulting centers are needed to run {\sf ellipse},  where we kept
the ellipticity and position angle parameters fixed at the values
measured by eye (see Figure \ref{7331g_dgal}).

In order to measure the surface brightness as a function of
galactocentric distance along the semi major axis, we set up {\sf
ellipse} to fit a series of concentric isophotes to the stamp images
of the four candidates with the length of the semi major axis
gradually increasing with a linear step of a factor of 1.1 (see Figure
\ref{7331g_dgal}).

\begin{figure}[h]
\includegraphics[width=3.425in]{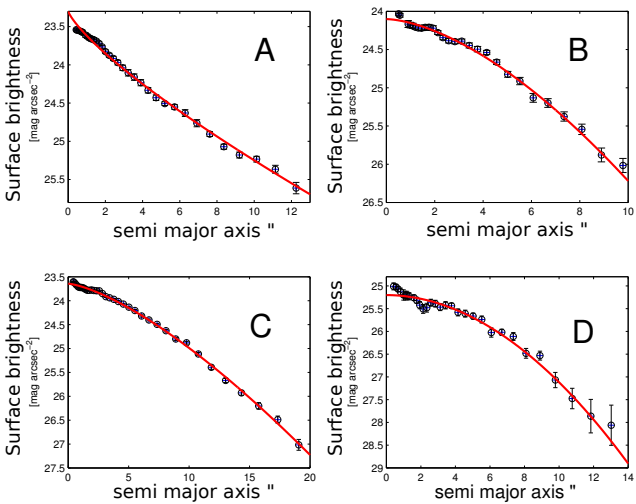}
\caption{Surface brightness profile fits to the dwarf galaxy 
candidates of NGC\,7331 in the SDSS $r$-band.
}
\label{sbp}       
\end{figure}

The surface brightness profiles (SBPs) derived in the $r$-band are
shown in Figure \ref{sbp} with open circles and the error bars
computed by {\sf ellipse}. The maximum length of their semi major axis
is defined by the corresponding surface brightness being just $1
\sigma$ above the background.  To characterize the structural
parameters of the four candidates, we fitted their SBPs with the 
S\'ersic profile \citep{Sersic1968Atlas}: 

\begin{equation}
 \mu (r) = \mu _{0} + 1.086 * (r/h)^{1/n} 
\end{equation}
\citep{Cellone1999A&A345}
where $\mu$ is the surface brightness, $\mu_{0}$ is the central surface
brightness, $r$ is the distance from the galaxy center along the semi major axis, 
$h$ is the scale length and $n$ is the S\'ersic index.
We cut each SBP at the semi major axis whose corresponding  
surface brightness is just 3$\sigma$ (1$\sigma$ for NGC\,7331\,D because of its low surface brightness) 
above the background
in order to minimize the uncertainty on the best fitting S\'ersic profile.
We used a non linear least-squares fit
with the Levenberg-Marquardt algorithm. 
The best fitting S\'ersic profiles are plotted in
Figure \ref{sbp} with a solid line, and their parameters ($n$,
$\mu_{0}$, $h_{r}$) are shown in Table \ref{dwarfdata}. 

The S\'{e}rsic indices $n$ obtained for the four candidates range from
0.5 to 1.3, and are in good agreement with those measured by
\citet{Chiboucas2009AJ137} for the dSphs and dIrrs in the M81 group
(0.21 $\lesssim$ $n$ $\lesssim$ 1.0).  The estimated $\mu_{0}$ of the
four candidates varies between 23 and 25~mag~arcsec$^{-2}$ and is
consistent with the central surface brightness of the dwarfs in the
M81 group as derived by \citet{Chiboucas2009AJ137}.

We used the SBPs to determine the semi major axes $r_{1\sigma}$ and
$r_{3\sigma}$ at which $\mu$ is just 1$\sigma$ and $3\sigma$,
respectively, above the background.  For our data, $r_{1\sigma}$ and
$r_{3\sigma}$  are smaller in the $g$-band, therefore we used their
values in the $g$-band ($5.5'' < r_{1\sigma} < 11.0''$ and $2.5'' <
r_{3\sigma} < 9.0$) to compute the integrated ($g-r$)$_{0}$ colors of
each candidate. These colors are  reported in Table \ref{dwarfdata}.

For each candidate except NGC\,7331\,A, the colors derived for the 1$\sigma$
and 3$\sigma$ thresholds are very similar; in the case of
NGC\,7331\,A the color for the 3$\sigma$ threshold is bluer than that
for 1$\sigma$. This indicates that this galaxy is bluer in its
center, perhaps akin to what is observed in some of the more
massive ``blue-core dwarf ellipticals'' \citep[dE(bc) as classified
by][]{Lisker2006AJ132_2} in the Virgo cluster.  For these dE(bc) galaxies,
\citet{Lisker2006AJ132_2} were able to show spectroscopically
that they did experience star formation in their centers less than a
Gyr ago.  Similar examples -- again among the more luminous dEs -- are
also known from the Local Group, where the close M31 dE companions
NGC\,185 \citep[e.g.,][]{Hodge1963AJ68,Butler2005AJ129} and NGC\,205
\citep[e.g.,][]{Cappellari1999ApJ515, Monaco2009A&A502} show recent star
formation in their centers.  NGC\,7331\,A may be a lower-luminosity
nucleated dSph counterpart of these systems.  Note that some Galactic
dSph galaxies, in particular Fornax, also show relatively recent star
formation that occurred just a few 100 Myr ago, though at the current
time they are no longer active \citep{Grebel1999IAUS192,Saviane2000A&A355}.

The $(g-r)_{0}$ colors of the four dSph candidates around NGC\,7331
range from 0.55 to 0.75 mag.  In order to compare them with the dwarfs
in the Local Group, we used the color transformation equations from
\citet{Jordi2006A&A460} for Population II stars to translate the
($g-r$)$_{0}$ colors of the four candidates into ($B-V$) colors, i.e.,
we apply the transformation relation $(B-V)=0.918\cdot(g-r)+0.224$.
The resulting ($B-V$) values vary between 0.7 and 0.9~mag and are
consistent with the ($B-V$) colors measured for LGS\,3, And\,I and
Leo\,I in the Local Group \citep[][]{Mateo1998ARA&A}.

\subsection{Dwarf Galaxy Candidates in the SDSS}

We checked that the four dwarf galaxy candidates found in our WIYN
data are also visible in the SDSS images available for NGC\,7331. We
were able to recover them, although they have a much lower S/N 
in the SDSS data than in our own data.  With a detection limit
of $\mu_{0}(g)\approx 27.0$~mag~arcsec$^{-2}$ the WIYN data are
slightly deeper than the SDSS in the individual filters.

According to the NASA Extragalactic Database (NED\footnote{URL:
http://ned.ipac.caltech.edu}), NGC\,7331 has the same morphological 
type as M31, yet is more than 1 mag brighter in the optical
($V$-band) and in the near-infrared ($H$-band).  It may thus be more
massive than M31 (see also Section 5.1).  
We then might expect it to be surrounded by a number of close 
dE companions similar to those found near M31. 
Interestingly we did not detect any bright dE across of the
field of view of our WIYN data.  We then decided to search for dE
candidates in the SDSS images covering the surroundings of NGC\,7331
out to a distance of 1~Mpc.  The depth of the SDSS imaging data in the
$g$-band reaches surface brightnesses as faint as
$\mu_{0}(g)\approx26.0$~mag~arcsec$^{-2}$ --
26.5~mag~arcsec$^{-2}$ \citep{Kniazev2004AJ127}.  Therefore dEs, which
have a typical $\mu_{r}$ between 21 and 24~mag~arcsec$^{-2}$ and a
($g-r$) color between 0.2 and 0.8~mag
\citep{Lisker2007ApJ660,Lisker2008AJ135}, should be easily detectable
in the SDSS images.

We retrieved from the SDSS DR7 photometric catalog
\citep{Abazajian2009ApJS182} all objects with a Petrosian radius
larger than $10''$, 0 $\lesssim$ ($g-r$) $\lesssim$ 1.5~mag, and
21~mag~arcsec$^{-2}$ $\lesssim \mu_{g} \lesssim$ 26~mag~arcsec$^{-2}$,
in order to sample the parameter space of dEs and dSphs. The limits in
color and surface brightness were chosen from the properties of dEs in
the Virgo Cluster \citep{Lisker2007ApJ660,Lisker2008AJ135}.  Smaller
Petrosian radii do not allow one to carry out a proper analysis of the
object's morphology.  A visual inspection of the selected objects
revealed no dE or bright dSph candidates.

\subsection{Analysis of the Stellar Stream}
\label{streamanalysis}

\begin{figure}[h]
\includegraphics[width=3.425in]{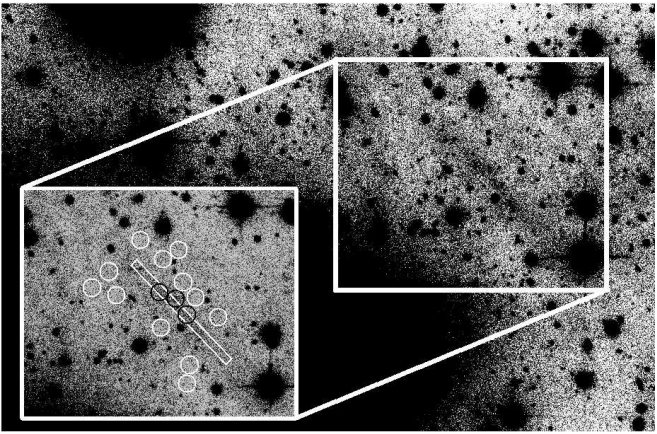}
\caption{Faint stellar stream near NGC\,7331. The inset shows the analyzed region.
 The black circles mark the brightest regions of
the stream where we measured the magnitudes and white circles represent the areas of
the local background estimation. The white rectangle indicates the region where the color profile 
was measured. }
\label{stream}       
\end{figure}

In the south eastern outer regions of NGC\,7331, we find hints of a
very low surface brightness structure in our WIYN images that we
interpret as a stellar stream (Figure \ref{stream}).  We measured the
length of the stream with the SAOImage DS9 tool `projection' by
placing a rectangular aperture,  29$'' \times$ 85$''$, along the
stream in a combined $g$-and $r$-band image. This  gives the mean
counts along the stream.  This produced an average brightness profile
of the stream, and its length was estimated as the distance between
the two edges of the stream where the brightness is less than
1$\sigma$ above the background.  

\begin{figure}[h]
\includegraphics[width=3.425in]{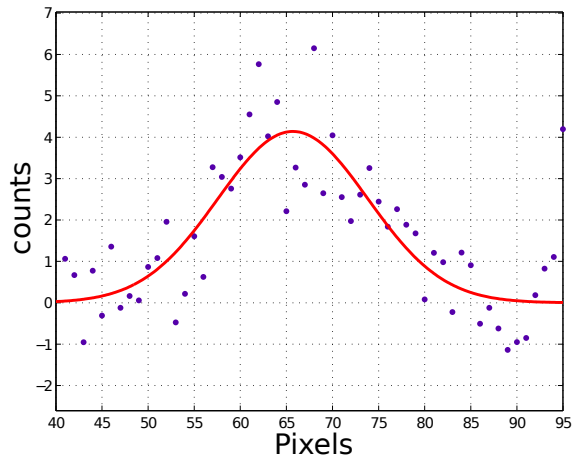}
\caption{Gaussian fit to the width of stream. For the fit, we used
the stacked $g$- and $r$-band images.}
\label{streamwidth}       
\end{figure}

Similarly, we determined the width of the stream with a box of 29$''
\times$ 65$''$ placed across the stream in a region free of bright
stars.  The resulting brightness profile was fitted with a Gaussian
function (see Figure \ref{streamwidth}). Its FWHM was used to define the width of the stream.  In
this manner we obtained a length of 5 kpc and a width of 802 pc for
the stream (assuming the same distance modulus as for NGC\,7331).

We defined a set of twelve circular apertures, 10$''$ in diameter, to
estimate an average background in the surroundings of the stream, and
three apertures of the same size centered on the brightest regions of
the stream to derive its surface brightness. After correcting for the
local background, we obtained $\mu_{g,0}=(26.88 \pm 0.11)$
mag~arcsec$^{-2}$, $\mu_{r,0}=(26.63 \pm 0.10)$ mag~arcsec$^{-2}$, and
$(g-r)_{0} = (0.16  \pm 0.16)$~mag, at an S/N of about 7.  We discuss
the possible origin of the stellar stream in Section 5.


\begin{figure}[h]
\includegraphics[width=3.425in]{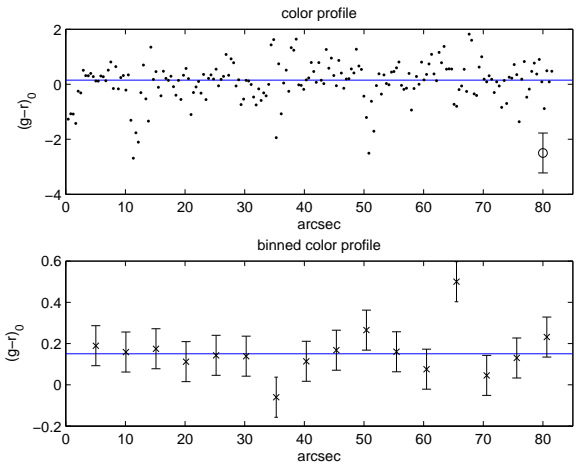}
\caption{Color profile of the stream. The upper plot shows the individual color estimates per pixel.
The bottom plot is binned with a bin size of 13 pixels. In both panels, the line represents the calculated mean.
Within the 1$\sigma$ errors no color gradient can be detected.}
\label{colorprof}       
\end{figure}

Since many dwarf galaxies in groups and clusters show color
gradients \citep[e.g.,][]{Kormendy1989ARA&A27,Chaboyer1994ESOC49,
Harbeck2001AJ122}, it is interesting to check whether the stream also
shows such a color variation. In order to do so we used {\sf
pvector} on an area of 354~kpc $\times 5$~kpc aligned with the stream
and extracted the average surface brightness of the stream in each
filter (see Figure \ref{stream}).  The resulting color profile is
plotted in the top panel of Figure \ref{colorprof}, where the solid
line represents the mean color obtained after a 1~$\sigma$ clipping of
the data.  We also binned the color profile with a bin size of 13
pixels in order to improve the S/N; the result is shown in the bottom
panel in Figure \ref{colorprof}, where the solid line traces again the
average color across the stream ($0.15 \pm 0.06$~mag). This
is in good agreement with the color previously obtained with aperture
photometry.  Integrating the SBP of the stream
we derived $g_{0}=(20.03 \pm 0.46)$~mag and $r_{0}=(19.87 \pm
0.69)$~mag corresponding to $M_{r}=(-10.83 \pm 0.76)$~mag at the
distance of NGC 7331.

We also checked the SDSS images in the $g$- ,$r$-, and
$i$-bands, but we were not able to clearly detect the stream.  Also the stacked
$gri$-band image does not show a recognizable stream, indicating
that the SDSS data are not deep enough and do not provide a high
enough S/N to study the stream.

In addition to the stream, another faint feature -- a plume -- can be
detected in the southwest outer regions of NGC\,7331 (see Figure
\ref{7331hc}).  We performed aperture  photometry (three apertures
with $r=5''$) of the plume and derived average surface brightnesses of
these apertures of $\mu_{g_{0}}=(26.29 \pm 0.46)$~mag and
$\mu_{r_{0}}=(25.79 \pm 0.30)$~mag with a resulting color of
$(g-r)_{0}=0.50 \pm 0.17$.  The errors correspond to the fluctuations
of the calculated magnitudes in the different apertures. In its
surface brightness this feature is similar to the northern spur that
can be seen around M31 \citep{Ferguson2002AJ124}. This feature could be
the result of another accretion event.  Given their relatively blue
colors, the stream and the plume cannot be mistaken for
Galactic cirrus clouds, because, as discussed in Section \ref{obsgg},
the latter have a typical $g-r$ color between 1.3 and 2~mag.

\section{Discussion}
\label{discussion}

Hierarchical models of galaxy evolution \citep[e.g.,][]{white1991ApJ}
suggest that the halos of giant galaxies contain large numbers of
low-mass dark matter halos.  If populated by baryonic matter as well,
these halos may be observable as dwarf galaxies.  It is well
established, however, that even in the case of the MW the number of
known dwarf companions is less than the predicted number of satellite
halos \citep[e.g.,][]{Moore1999ApJ524,Kravtsov2010AdvAst}.  While
observations of the Local Group and other nearby groups
\citep[e.g.,][]{Karachentsev2005AJ129} show a high number of dwarf
galaxy companions and some tidal interactions
\citep[e.g.,][]{Miskolczi2011A&A536}, the numbers are consistently
lower than the model predictions. This missing satellite problem
indicates that on the mass scale of galaxies, observations do not
reveal luminous versions of the expected substructures. 
 
The possibility therefore exists that only a minority of low-mass
satellite dark matter halos contain stars
\citep[e.g.,][]{Madau2008ApJ679}. An interesting question is how
common such observable luminous structures are, and whether they
follow well-defined patterns associated with either host galaxy or
galaxy group properties in the nearby universe. Thus we selected
nearby galaxy groups for study in GGADDS as these provide the most
common environments within which sub-halos should exist in the 
present-day universe.

\subsection{The NGC\,7331 Galaxy Group}

\begin{figure}[h]
\includegraphics[width=3.425in]{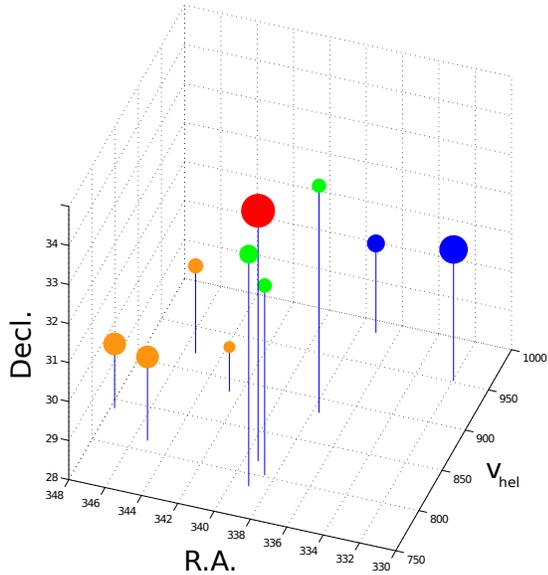}
\caption{Overview of the NGC\,7331 group. The red dot stands for 
NGC\,7331. The blue dots mark
the subgroup around NGC\,7217, the green ones show other spiral
and irregular galaxies belonging to the NGC\,7331 subgroup, 
and the yellow dots indicate the NGC\,7457 subgroup
(the dot size represents 
the luminosity of each galaxy).  The right ascension (R.A.) and the 
declination (Decl.) are given in degrees.  The heliocentric
velocity, $v_{hel}$ is given in km~s$^{-1}$.}

\label{groupstructure}       
\end{figure}

NGC\,7331 is a member of a filamentary group of galaxies (see
Figure~1).  It contains two giant spirals, NGC\,7217 and NGC\,7331, as
well as the luminous S0 galaxy NGC\,7457.  If we consider the virial
radii of typical galaxy groups to be $R_{vir} \leq1.3$~Mpc
\citep{Karachentsev2005AJ129}, then the NGC\,7331 group appears to
consist of three subgroups: (1) the relatively empty zone
surrounding the SAb galaxy NGC\,7217; (2) a subsystem
containing the late-type galaxies NGC\,7320, UGC\,12060, and
UGC\,12082 associated with NGC\,7331; and (3) a third
grouping around the S0 galaxy NGC\,7457, which is accompanied by a dwarf
S0/dE.
The NGC\,7457 subgroup is noteworthy in containing early-type disk
galaxies, suggesting that it is dynamically more evolved. In contrast
to that the NGC\,7331 subgroup is dominated by gas-rich late-type
galaxies, indicating that it is likely to be in an earlier dynamical
evolutionary phase (see Figure \ref{groupstructure}).

 Overall, the NGC\,7331 subgroup therefore resembles the
loose groupings of late-type giant galaxies found in the Sculptor
group or the Canes Venatici cloud \citep{Jerjen1998AJ116,
Karachentsev2003AAp404, Karachentsev2003A&A398,
Karachentsev2005AJ129}.  \citet{Karachentsev2005AJ129} finds that the
M94 group and the Sculptor group are not in a state of dynamical
equilibrium, a situation that likely applies to the overall NGC\,7331
group.  In comparison the Local Group, with its two giant spirals
separated by $\sim 0.8$~Mpc, is a denser and likely more massive galaxy
group  than the NGC\,7331 subgroup.  Such a more massive and 
richer galaxy group should have a shorter crossing time, and offer more
opportunities for internal interactions within the typical
group radius of approximately 1 Mpc. 

\subsection{The NGC\,7331 Dwarf Sample Compared to M31 and Other
Nearby Galaxies}

Our study of the region surrounding NGC\,7331 reveals four candidate
dwarf companions. Three of them have $M_r \leq -$13~mag and are
located within a projected radius of 160~kpc from NGC\,7331. As this
corresponds to the same region covered by the main Pan-Andromeda
Archaeological Survey \citep{Richardson2011ApJ732}, it is interesting
to compare results.  With our sensitivity we would detect the
compact elliptical satellite M32 and all three bright dEs around M31
(i.e., NGC\,147, NGC\,185, and NGC\,205). Obviously any counterparts
of such galaxies are missing from our survey zone surrounding
NGC\,7331.  We would also detect at least two of the brighter M31 dSph
companions, And~I and And~II; thus our data are suggestive of roughly
comparable luminous dSph populations in the two systems. This
result is re-enforced by our finding that the NGC\,7331 dwarf galaxy
candidates have colors and structures of typical dSph galaxies. 

As the  absolute $H$-band magnitude for NGC\,7331
\citep{Aaronson1977PhDT} is $\approx$ 1~mag brighter than that for
M31, it is likely that M31 has less stellar mass than NGC\,7331.  In
addition the bulge-to-disk luminosity ratio for NGC\,7331 is B/D
$\approx$1  \citep{Bottema1999AA348} versus B/D $\approx$ 0.5 for M31
\citep{Geehan2006MNRAS.366}.  Thus if the dSph population were tied to
bulge luminosity, we might expect richer dwarf populations in
NGC\,7331 as compared to M31. Yet apparently this is not the case, at
least within our magnitude limit and with $R\leq$160~kpc.  The
deficiency of early-type dwarf satellites associated with NGC\,7331,
however, is more pronounced if we consider the lack of equivalents to
the three classical, luminous dE M31 companions.  If we compare the
stellar masses of more luminous early-type dwarf companions, excluding
the anomalous M32 system, then the detected stellar mass in the form
of candidate satellite galaxies around NGC\,7331 is about 2\% of that
found in the same radius and absolute magnitude limit around M31.  It
thus seems possible that the presence of dE-type satellites could be
linked to some parameter other than the (stellar) mass of the host
galaxy, such as group density, as suggested by earlier studies
\citep[e.g., ][]{Ferguson1991AJ101}.

We conclude that unless dE companions to NGC\,7331 have been missed at
larger radii, where they should have been detected as candidate dwarfs
in the SDSS images, NGC\,7331 has substantially less stellar material
in the form of satellites than the less massive M31 system.
Moreover, NGC\,7331 lacks luminous nearby companions like the
Magellanic Clouds around the MW.  Such irregular companions, 
however, are
generally found to be rare \citep[e.g.,][]{Guo2011MNRAS417}.

\begin{figure}[h]
\includegraphics[width=3.425in]{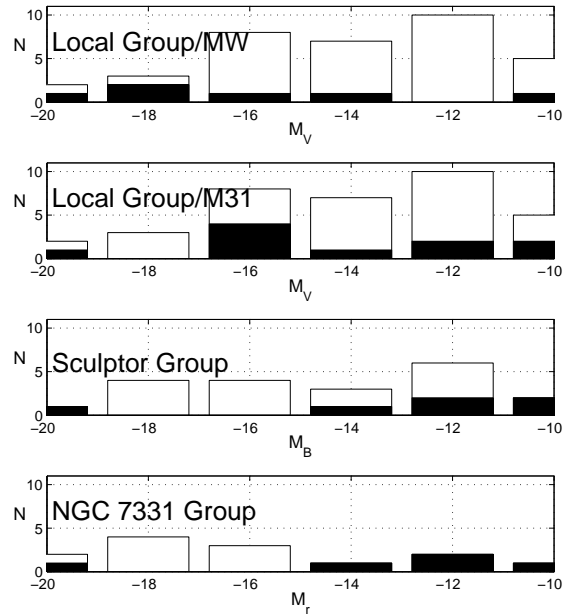}
\caption{Luminosity function of the NGC\,7331 subgroup (bottom panel) 
compared to the Milky Way and M31 subgroups in the Local Group (upper 
two panels) and the NGC\,253 subgroup in the Sculptor group (third panel). 
The white (unfilled) histogram bars show the total number of detected 
member galaxies in a given group.  The black (filled) histogram bars
in the panels indicate what we would measure if we take into account 
only the likely dwarf
satellites of one massive disk galaxy (the Milky Way and M31 for the 
Local Group and NGC 253 for Sculptor) to permit a direct comparison
with NGC\,7331. 
}
\label{lumfct}       
\end{figure}

The statistics for dSphs with $M_r \leq -$14 are not well
constrained. Our few detections only sample the counterparts of the most
luminous counterparts of the ``classical'' dSphs in the Local Group.
When comparing NGC\,7331 and M31 in this respect, it seems possible
that similarly numerous populations of such dSph satellites may exist
around both spiral galaxies.  The differences and similarities between
the populations of inner satellite galaxies around NGC~7331 as well as
in the Sculptor group and in the Local Group are further illustrated in
Figure \ref{lumfct}.

\subsection{The Stellar Stream}

Stellar streams extending out of the planes of disk galaxies are well
established as indicators of past interactions.  Given the low density
of the NGC~7331 galaxy group, the presence of stellar streams in
NGC~7331 then is somewhat surprising.  We focus our discussion on the
western stream, but note that additional features are seen to the
southwest of the galaxy.  As the streams appear to be discontinuous
and given that the estimated color of the southwestern feature is
different from that of the western stream, we cannot tell whether
these are two parts of one single stream or constitute two separate
structures. 

If the western stellar stream is tidal debris, then the galaxy
producing it might be detectable. A system with the luminosity of $M_r
= -10.8$~mag that we measured in the western stream segment located
away from the obscuring main body of NGC\,7331 would be included in
our sample, but obviously was not found.  The blue color of this
stream segment, $(g-r)_0 = 0.15$~mag, suggests that it might
have been produced by a star-forming galaxy (possibly a dIrr or a
dIrr/dSph transition-type galaxy), even though no H\,{\sc i} is seen
today to be associated with the stream. The narrow projected
width of the NGC~7331 stream indicates that such a progenitor would
have had a low internal stellar velocity dispersion
\citep{Johnston2008ApJ}.  We thus speculate that the accreted galaxy
could have been a low-mass dIrr or dIrr/dSph transition-type galaxy,
consistent with the modest luminosity and low surface brightness
($\mu_{r,0} = 26.6\pm0.1$~mag~arcsec$^{-2}$) of the western stream
segment. 

According to the color transformations of \citet{Fukugita1995PASP107}
[$(B-V) = (g-r) + 0.07$, computed for irregular galaxies], the ($g-r$)
color of the stream translates into $(B-V)=0.23$~mag, comparable with
the colors of the bluest dIrrs in the Local Group \citep[e.g., DDO210,
Leo\,A, Sex\,A, SagDIG from][]{Mateo1998ARA&A}.  The observed
blue color could alternatively be produced by old, extremely
metal-poor stellar populations, which in addition may possess very
blue horizontal branches.  However, a comparison with the integrated
$B-V$ colors of Galactic globular clusters from \citet[][2010
edition]{Harris1996AJ112} does not reveal any comparable objects even at
low metallicities.  The ultra-compact dSph galaxies detected around
the MW in recent years \citep[e.g.,][]{Zucker2006ApJL650,
Belokurov2006ApJL647} tend to be even more metal-poor than globular
clusters, but contain so few stars that they would not leave a
recognizable stellar stream with the surface brightness of our stream
candidate.  The blue color of the stream could also indicate tidally
triggered star formation caused by the disruption event if indeed the
progenitor was a gas-rich dIrr galaxy.  Such young stellar debris was
found, for instance, in the halo of the peculiar elliptical galaxy
NGC\,5128 (Centaurus\,A), and its A-star colors resemble those of our 
candidate \citep{Peng2002AJ124}.

The timescale since formation of the stream is uncertain.  If the
NGC\,7331 stream is the result of an interaction with a 
gas-rich dwarf contributing either young blue stars or undergoing
tidally triggered star formation, which might suggest a fairly recent
interaction a few Myr ago \citep[see][]{Peng2002AJ124}.  In the Local
Group, stars with ages possibly as young as 800 Myr have been detected
in the central regions of the Sgr dSph galaxy \citep{Siegel2007ApJ667},
which is merging with the MW.  These stars formed long {\em
after} Sgr began disrupting, but represent only a tiny fraction of the
overall stellar populations of Sgr.  In any case, interactions
involving or leading to recent star formation may not be unusual.
Alternatively, such a blue feature might, in principle, form from gas
accreted from an intra-group medium.  But the observed narrowness of
the stream makes this scenario questionable.  While star formation has
been observed in tidal debris \citep{Makarova2002AAp396,deMello2008AJ135,
Werk2008ApJ678}, to our knowledge it is not seen in the cases where
{\em diffuse} H\,{\sc i} is thought to be accreted onto galaxies
\citep[e.g.,][]{Sancis2008AARv15}, and therefore is a less likely
possibility.

\section{Conclusions}
\label{conclusions}

We present the first results from GGADDS in the form of a survey for
candidate tidal features and dwarf galaxies in a 160~kpc radius region
surrounding the giant SAb spiral NGC\,7331, located at an approximate
distance of 14.2~Mpc.   NGC\,7331 is the primary member of a subgroup
within a larger filamentary and relatively diffuse galaxy group of the 
same name. Our deep
$g$- and $r$-band images reveal four dwarf galaxy candidates that have
colors and structures consistent with dSph satellites close to
NGC\,7331.  No examples of more luminous dE-like companions are found.
Moreover, a faint western stellar stream is detected, which has a blue
color and low surface brightness, possibly indicative of having had a
star-forming, low-mass dIrr/dSph transition-type galaxy or a gas-rich
dIrr undergoing tidally triggered star formation as its progenitor.
There is also a more diffuse and less well-defined feature to the
southwest, but it is unclear whether it is connected with the western
stream.

Even though NGC\,7331 is a giant spiral with higher total and bulge
luminosities than M31, it apparently lacks M31's retinue of
comparatively luminous dE satellites, as well as something like its
compact elliptical companion M32.  The absence of such luminous,
massive dE companions means that the stellar mass in the form of inner
satellites is 2\% or less of that around M31. 

On the other hand, our few candidate dSph detections are consistent
with NGC\,7331 and M31 having comparable populations of (classical)
dSph companions.  The results of our study therefore agree with
surveys showing highly variable numbers of moderate-luminosity
satellites of giant galaxies, while suggesting that dSph satellites
may be more closely tied to the properties of the main galaxy.
The source of the large variance in spheroidal satellite
stellar masses in galaxy groups remains an interesting question and a
potential clue to the origin of these galaxies.  

Given the paucity of neighboring galaxies, the detection of what
appears to be a bluish stellar tidal debris stream is unexpected.  The
most straightforward explanation is that NGC\,7331 interacted with and
at least partially disrupted a star-forming low-mass transition-type
dwarf, or that the accretion of a gas-rich dIrr triggered star
formation.  One of the four dSph candidates discovered in our study
shows a blue core similar to more massive, nucleated dEs with recent
central star formation found in the Virgo cluster, or similar to
transition-type dwarfs with residual centralized star formation
activity as observed in the Local Group.  Deeper optical and H\,{\sc
i} observations would be useful in order to obtain a better
understanding of the dwarfs and stream(s) around NGC\,7331.  In
low-density groups like the diffuse, extended NGC\,7331 group with its
many late-type galaxies, interactions with gas-rich dwarfs may be more
common than in higher-density environments.

\section*{Acknowledgements}

We thank the anonymous referee for useful comments, which helped to improve this paper.
We acknowledge the WIYN Consortium for providing telescope time for this
project and for maintaining the WIYN 0.9-m Observatory, along with
Hillary Mathis for her support of the Observatory.  We are also
grateful to the Heidelberg Graduate School of Fundamental Physics (DFG
grant No. GSC 129/1), which provided funding for a research visit
to Madison, and to the University of Wisconsin-Madison Graduate School
for partial support of this research.  This research made use of the
HYPERLEDA database \citep{Paturel2003A&A412} and the NASA/IPAC
Extragalactic Database (NED) which is operated by the Jet Propulsion
Laboratory, California Institute of Technology, under contract with
the National Aeronautics and Space Administration. 

\bibliography{bibjo.bib}
\bibliographystyle{apj}

\end{document}